\documentclass[11pt,a4paper]{article}
\usepackage{jheppub,epsfig,multicol,bbm,latexsym,graphicx,subfigure,color}

\title{Constraints on dark matter annihilation from the Event Horizon Telescope Observations of M87$^\star$}

\author{Guan-Wen Yuan$^{a,b}$, Zhan-Fang Chen$^{a,b}$\footnote{Corresponding author}, Zhao-Qiang Shen$^{a,b}$, Wen-Qing Guo$^{a,b}$, Ran Ding$^{c}$\footnote{Corresponding author}, Xiaoyuan Huang$^{a,b}$, and Qiang Yuan$^{a,b}$}

\affiliation{
$^a$Key Laboratory of Dark Matter and Space Astronomy, Purple Mountain Observatory, Chinese Academy of Sciences, Nanjing 210023, China\\
$^b$School of Astronomy and Space Science, University of Science and Technology of China, Hefei 230026, China\\
$^c$School of physics and optoelectronics engineering, Anhui University,
Hefei 230601, China}
\emailAdd{yuangw@pmo.ac.cn}
\emailAdd{chenzf@pmo.ac.cn}
\emailAdd{zqshen@pmo.ac.cn}
\emailAdd{guowq@pmo.ac.cn}
\emailAdd{dingran@mail.nankai.edu.cn}
\emailAdd{xyhuang@pmo.ac.cn}
\emailAdd{yuanq@pmo.ac.cn}

\abstract{The fast developments of radio astronomy open a new window to explore the properties of Dark Matter (DM). The recent direct imaging of the supermassive black hole (SMBH) at the center of M87 radio galaxy by the Event Horizon Telescope (EHT) collaboration is expected to be very useful to search for possible new physics. In this work, we illustrate that such results can be used to detect the possible synchrotron radiation signature produced by DM annihilation from the innermost region of the SMBH. Assuming the existence of a spike DM density profile, we obtain the flux density due to DM annihilation induced electrons and positrons, and derive new limits on the DM annihilation cross section via the comparison with the EHT integral flux density at 230 GHz. Our results show that the parameter space can be probed by the EHT observations is largely complementary to other experiments. For DM with typical mass regions of being weakly interacting massive particles, the annihilation cross section several orders of magnitude below the thermal production level can be excluded by the EHT observations under the density spike assumption. Future EHT observations may further improve the sensitivity on the DM searches, and may also provide a unique opportunity to test the interplay between DM and the SMBH.}

\arxivnumber{2106.05901}

\begin{document}
\maketitle
\flushbottom

\section{Introduction}

The nature of Dark Matter (DM) is one of the biggest problems in modern physics and cosmology \cite{Rubin:1980zd, Clowe:2006eq, Larson:2010gs}. Quite a lot of particle models beyond the standard model were proposed in literature \cite{Jungman:1995df, Bertone:2004pz, Feng:2010gw, Roszkowski:2017nbc}. Some classical candidates are well-motivated, such as weakly interacting massive particles (WIMPs) \cite{Roszkowski:2017nbc, Liu:2017drf}, axions or axion-like particles \cite{Peccei:1977hh, Weinberg:1977ma, Davoudiasl:2019nlo, Yuan:2020xui}, and dark photons \cite{An:2020jmf, Caputo:2021eaa}. Meanwhile, many detecting methods and technologies have also been developed, such as detecting the feasible energy missing in particle collisions \cite{Slatyer:2017sev}, observing the scattering signals between DM and detector particles \cite{Liu:2017drf,Aprile:2020tmw}, and measuring the DM induced products in cosmic rays \cite{Porter:2011nv, Ge:2020yuf, Cao:2020bwd, Chen:2020gcl}. Among these candidates, WIMPs with typical masses from sub-GeV to multi-TeV are most widely studied, due partly to the so-called ``WIMP miracle'' the corect relic density of DM can be naturally obtained in this model.

WIMPs can undergo annihilation or decay, yielding standard model elementary particles by multi-channels, such as electrons and positrons. The electrons and positrons can imprint in the electromagnetic spectrum due to radiative processes such as synchrotron, inverse Compton scattering, and/or bremsstrahlung radiation~\cite{Profumo:2010ya, McDaniel:2017ppt, McDaniel:2018vam}.
There have been considerable efforts to study multi-wavelength electromagnetic emission from DM annihilation in a variety of galaxy clusters and dwarf galaxies, e.g., Refs.~\cite{Yuan:2009yy, Huang:2011xr, Bringmann:2012vr, Bergstrom:2013jra, Ackermann:2015zua, Slatyer:2015jla, Li:2015kag, Abdallah:2016ygi, Li:2018kgy,Cang:2020exa}. A sizable parameter space of WIMPs was excluded by these observations.

The center of a galaxy is expected to be a site with potentially maximum DM concentration, although the detailed DM density profile remains uncertain in the galaxy center due to the effect from the baryon interaction and/or the accretion of the SMBH \cite{Ferrarese:2004qr,Vasiliev:2008uz,Gondolo:1999ef, Gnedin:2003rj, Fields:2014pia}.
%There are some DM density profiles have been widely discussed and applied, such as solitonic core for ultralight DM \cite{Schive:2014dra, Schive:2014hza}, spike profile for cold DM \cite{Gondolo:1999ef, Gnedin:2003rj, Fields:2014pia}.
The previous observations with poor spatial resolutions are difficult to probe the environment parameters close to the black hole, and hence increase the uncertainties of the signal prediction.
With the development of technologies of the Very Long Baseline Interferometry (VLBI) \cite{Middelberg:2008qc}, the Event Horizon Telescope (EHT) program, a global network of millimeter and sub-millimeter observational facilities, has been established successfully and has imaged the SMBH M87$^\star$ to the event horizon scale with extremely high angular resolution \cite{Akiyama:2019cqa, Akiyama:2019bqs, Akiyama:2019fyp, Akiyama:2019eap}. Besides crucial tests of the physical laws in the strong gravity field, the EHT observations open a new window to study broad types of new physics models including the particle nature of DM \cite{Davoudiasl:2019nlo, Chen:2019fsq, Chen:2021lvo}. The EHT observations enable one to scrutinize the physical and astrophysical processes just surrounding the SMBH, and can thus provide a unique probe of the DM interplay with the black hole.

In this work, we search for radio emission signature induced by the DM annihilation at subparsec scales using the EHT M87$^\star$ results~\cite{Akiyama:2019cqa, Akiyama:2019bqs}. For the DM density distribution in the center of the galaxy, a spike profile is expected due to the adiabatic accretion of the SMBH \cite{Gondolo:1999ef, Ullio:2001fb, Gnedin:2003rj}, and the number of electrons and positrons annihilated by DM would be enhanced several orders.
We will discuss the consequence of the possible radio emission via the synchrotron radiation by high-energy electrons and positrons in the magnetized accretion disk of M87$^\star$. Particularly, we find new properties of the relation between the WIMP annihilation cross section and the synchrotron fluxes, due to the annihilation saturation of the density near the horizon. Consequently a finite range on the cross section and mass plane is able to be probed by the data.
%We derive the synchrotron radiation by solving the propagation equation of DM-induced electron and positrons in presence of advection, which describe injection particles inflow toward the center of SMBH and enhance the electrons/positrons energy spectrum several orders, and obtain the strongest bounds on annihilation cross-section at different channels (such as  $\mu^{+}\mu^{-}$, $\tau^{+}\tau^{-}$, $b\bar{b}$ and $e^{+}e^{-}$).

The rest of this paper is organized as follows. In Sec.~\ref{spike}, we present the spatial distribution of DM around the SMBH, especially the DM spike profile. In Sec.~\ref{electron propagation around SMBH}, we review the propagation of electrons and positrons with the synchrotron cooling and advection effect, and then derive the limits on DM parameters using the EHT data. We conclude our work with some discussion in Sec.~\ref{summary}.

\section{Density profile of DM halo with SMBH}\label{spike}

We assume that the global DM halo density profile of M87 galaxy is described by the Navarro-Frenk-White (NFW) profile, which is expected to be a universal density profile for cold DM halo in hierarchically clustering Universes~\cite{Navarro:1995iw, Navarro:1996gj}. It is parameterized as
\begin{align}\label{NFW_profile}
\rho_{\mathrm{NFW}}(r)=\frac{\rho_0}{\left(r / r_{0}\right)\left(1+r / r_{0}\right)^{2}},
\end{align}
where $r_0$ is the scale radius and $\rho_0$ is the normalization constant. Eq.~(\ref{NFW_profile}) shows a $r^{-1}$ density cusp at the center.

However, as proposed by Gondolo and Silk~\cite{Gondolo:1999ef}, the adiabatic accretion of DM onto the SMBH in the galactic center would significantly enhance the DM density around it and form a spiky structure in the DM density distribution. Further studies showed that some dynamical effects, such as the mergers of black holes and the interactions with the baryonic matter, may shallow or even disrupt the spike \cite{Merritt:2002vj,Gnedin:2003rj,Bertone:2005hw}. Nevertheless, we assume that the DM spike had formed around M87$^\star$ and was survived from such dynamical effects. In this scenario, the spike density profile reads~\cite{Merritt:2002vj}:
%However, DM distribution in the micro-arcsec scales region of galaxies are poorly known at present and have been debate over two decades.
%Astronomical observation favour flat profiles, and numerical simulations support steeper profiles, which characterized by a slope parameter, as the DM density will be significantly enhenced by the adiabatic growth of SMBH at %Galactic Center and form a spike structure~\cite{Gondolo:1999ef, Ullio:2001fb, Gnedin:2003rj}.
\begin{equation}\label{spike_profile}
\rho_{\rm sp}(r)=\rho_0\left(\frac{R_{\rm sp}}{r_0}\right)^{-\delta}\left(\frac{R_{\rm sp}}{r}\right)^{\delta_{\rm sp}},
\end{equation}
where $\delta$ is the slope of the original density profile, and $\delta_{\rm sp}=(9-2\delta)/(4-\delta)$. We take $\delta = 1$ to match the NFW profile. $R_{\rm sp}$ is the spike radius, which is given as
\begin{equation}\label{spike_radius}
R_{\rm sp}=\alpha_\delta r_0 \left(\frac{M_{\rm BH}}{\rho_0 r_0^3}\right)^{\frac{1}{3-\delta}}.
\end{equation}
In the above equation, $M_{\rm BH}$ is the mass of black hole, $\alpha_{\delta}$ is a normalization coefficient which characterize the mass ratio of DM spike and black hole. Noticed that the depletion of DM density in the inner region of spike due to DM annihilations set a saturate DM density $\rho_{\rm sat}= m_\chi/\langle\sigma v\rangle t_{\rm BH}$. where $t_{\rm BH}$ is the age of the black hole, and $m_{\chi}$ and $\langle\sigma v\rangle$ are repectively the mass and thermally averaged annihilation cross section of DM particle. Using $\partial \rho_{\chi}/\partial t= - \langle\sigma v\rangle \rho_{\chi}^2/m_{\chi}$, one obtains spike density within the
region $R_{\rm Sch} \leq r<R_{\rm sp}$~\cite{Gondolo:1999ef}:
\begin{align}\label{spike_density}
\rho_\chi(r)=\frac{\rho_{\rm s p}(r) \rho_{\rm s a t}}{\rho_{\rm s p}(r)+\rho_{\rm sat}},
~\end{align}
where $R_{\rm Sch}$ is the Schwarzschild radius. Based on above analysis, when taking into account the effects of accretion of SMBH and DM annihilations, one should modify the NFW profile to the following piecewise distribution \cite{Aloisio:2004hy, Lacroix:2016qpq}:
\begin{align}
\label{profile}
\rho_\chi(r)=\left\{\begin{array}{ll}
0 & \qquad r<R_{\rm Sch},\\
\frac{\rho_{\rm s p}(r) \rho_{\rm s a t}}{\rho_{\rm s p}(r)+\rho_{\rm sat}} & \qquad R_{\rm Sch} \leq r<R_{\rm sp}, \\
\rho_{\mathrm{NFW}}(r)
& \qquad r \geq R_{\rm sp}.
\end{array}\right.
\end{align}

\begin{figure}[!htbp]
\begin{center}
\includegraphics[width=0.8\linewidth]{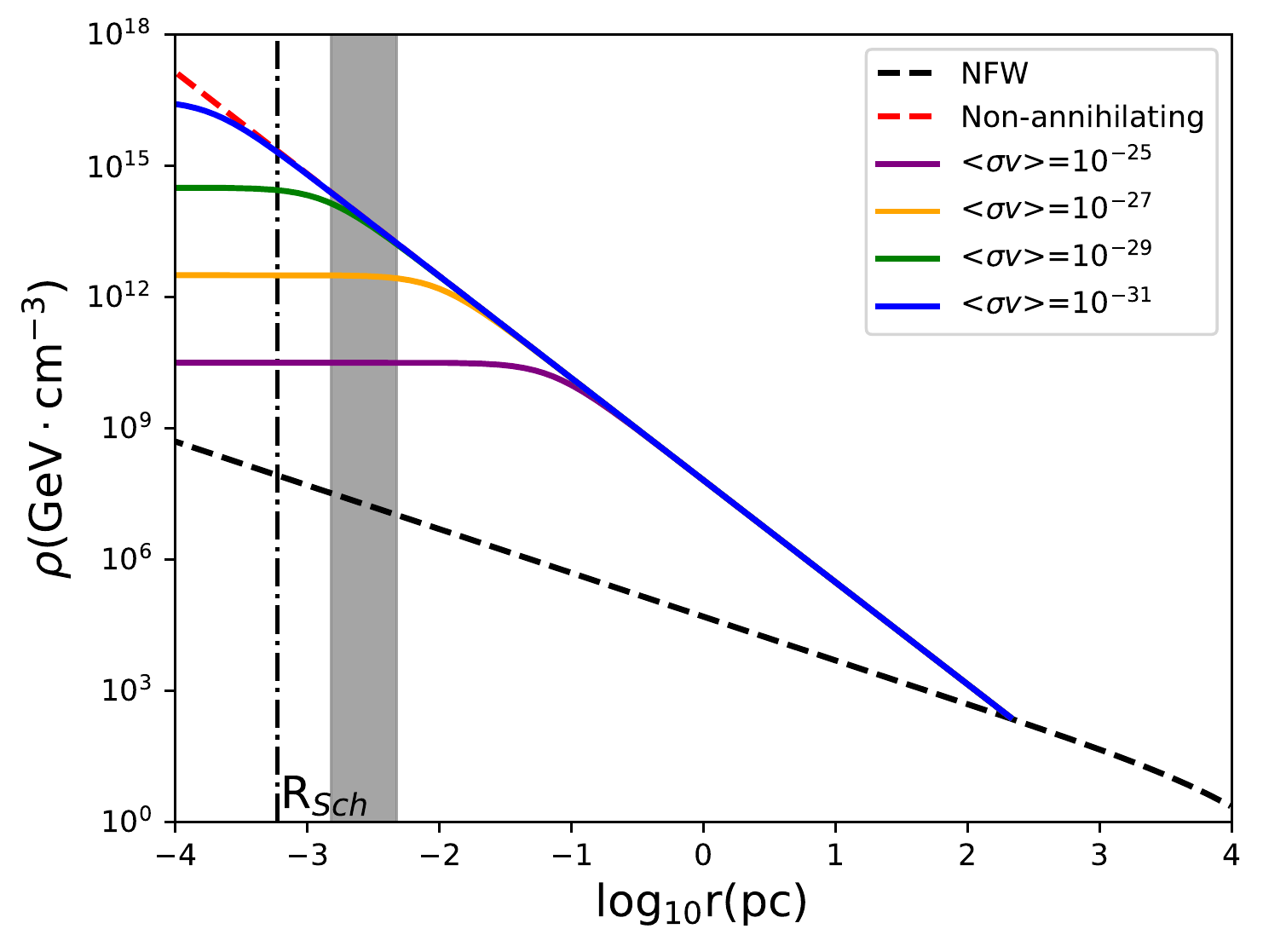}
\end{center}
\caption{Dark Matter Density Profile of M87$^\star$. Density distribution for an NFW halo profile (black dashed) with $\delta=1$ and the same halo with a spike in the central region for a non-annihilating DM (red dashed). Also shown are DM with $m_\chi=100$ GeV and thermally average annihilation cross section $\langle\sigma v\rangle= 10^{-31}, 10^{-29}, 10^{-27}$, and $10^{-25}~\mathrm{cm}^{3}~{\rm s}^{-1}$ (blue, green, orange and purple solid line), respectively. The dot-dashed black vertical line shows corresponding Schwarzschild radius of M87$^\star$, and the gray shadowed region corresponding to the areas covered by EHT observation. In this plot, the lifetime of SMBH is chosen with $10^{9}$ years and $R_{\rm sp}\simeq 220$ pc.}
\label{Fig.spike}
\end{figure}

According to the EHT results, the mass of M87$^\star$ is $6.4 \times 10^{9} ~M_{\odot}$, and the corresponding Schwarzschild radius is $R_{\rm Sch}=5.9\times 10^{-4}$~pc. Taking $\alpha_{\delta}=0.1$, we fix $r_0=20$~kpc for the halo (as for the Milky Way), and normalize $\rho_{0} \approx 2.5~\mathrm{GeV}~\mathrm{cm}^{-3}$ as Ref.\cite{Lacroix:2016qpq} done. Then, we could calculate $R_{\rm sp}\simeq 220$~pc throught Eq.(\ref{spike_radius}). Assuming the age of M87$^\star$ is $t_{\rm BH}=10^9$ yr and using Eq.~(\ref{profile}), we displayed the DM density profile of M87$^\star$ in Fig.~\ref{Fig.spike}. In this plot, the corresponding Schwarzschild radius of M87$^\star$ and the areas covered by EHT observations are also shown.

\section{Synchrotron emission induced by DM annihilations at the center of M87}\label{electron propagation around SMBH}

\subsection{Propagation of the Electrons and Positrons }

%The synchrotron emission through radiative losses involves charged particles, mainly electrons and positrons. Produced in WIMP pair annihilations, they propagate, losing and/or gaining energy. Thus, we need to compute the electron and positron spectra from the DM annihilation rate for deriving the intensity of DM-induced synchrotron radiation.
The distribution of electrons and positrons from the DM annihilation can been obtained via solving the propagation equation in the presence of synchrotron radiation and advection, which reads  \cite{Aloisio:2004hy, Regis:2008ij, Lacroix:2016qpq}
\begin{equation}
\label{propagation}
    -\frac{1}{r^{2}} \frac{\partial}{\partial r}\left[r^{2} D \frac{\partial f_e}{\partial r}\right]+v \frac{\partial f_e}{\partial r}-\frac{1}{3 r^{2}} \frac{\partial}{\partial r}\left(r^{2} v\right) p \frac{\partial f_e}{\partial p}+\frac{1}{p^{2}} \frac{\partial}{\partial p}\left(\dot{p} p^{2} f_e\right)=q(r, p)
\end{equation}
where $f_e(r,p)$ is the equilibrium distribution function of electrons and positrons in momentum space at radius $r$ and momentum $p$, assuming a steady-state. The first term on the left hand side describes the spatial diffusion, with $D(r,p)$ being the diffusion coefficient. The second and third terms are the advection current of accretion flow and the energy gain of electrons and positrons caused by adiabatic compression, with
$v(r)=-c\sqrt{R_{\rm Sch}/r}$ is the radial infall velocity of electrons and positrons in the accretion flow. Finally, the last term describe the energy loss due to radiative processes. We here consider the stationary state solution of the transport equation~(\ref{propagation}) in the spherical symmetry geometry.

The source function $q(r,p)$ is the DM annihilation rate in momentum space, which is related to the annihilation rate $Q(r,E)$ in energy space through following equation:
\begin{equation}
q(r,p) = \frac{c}{4\pi p^2}Q(r,E)=\frac{c}{4\pi p^2} \frac {\langle\sigma v\rangle\rho_\chi^2(r)}{2 m^2_{\chi}} \sum_i {\rm BR}_i\frac{d N^{\rm inj}_{e^\pm, i}}{d E}(E),
\end{equation}
where $\langle\sigma v\rangle$ is the thermally averaged annihilation cross-section, and $d N^{\rm inj}_{e^\pm,i}/dE$ is the $e^\pm$ injection spectrum through annihilation channels $i$ with branching ratios ${\rm BR}_i$. In this paper, we respectively consider annihilation channels $i=e^{+}e^{-},~\mu^{+}\mu^{-}, ~\tau^{+}\tau^{-},~b\bar{b}$ with $100\%$ branching ratios, and extract corresponding $e^\pm$ injection spectrum from the {\tt DarkSUSY} or {\tt PPPC} packages~\cite{Cirelli:2010xx}. The electron and positron energy spectrum is then written as
\begin{equation}\label{number density}
n_{e}(r, E)=\frac{4 \pi p^{2}}{c} f_{e}(r, p).
\end{equation}

In order to appropriately describe the propagation of electrons/positrons produced by DM annihilations with taking into account the influence of SMBH, one need to consider two propagation zones separately. Firstly, for $r$ outside the accretion radius $r_{\rm acc}$, the only energy loss mechanism for electron/positrons are radiative processes discussed above. While for $r\leq r_{\rm acc}$, one has more complicate picture due to the existence of
accretion flow towards central SMBH. As a consequence, at a given injection radius $R_{\rm inj}$, corresponding energy transfer of electron/positrons is governed by the competition between two physical processes, i.e., energy loss due to conventional radiative processes and the energy gain due to adiabatic compression along the plasma accretion flow. For the ultra-relativistic electrons/positrons with their radiative energy loss is dominated by synchrotron emission, transport equation~(\ref{propagation}) yields an integral analytic solution~\cite{Aloisio:2004hy}:
\begin{equation}\label{propagation_solution}
f_{e}(r, p)= \int_{r}^{r_{\mathrm{acc}}}\frac {Q_{i}\left(R_{\mathrm{inj}},p_{\mathrm{inj}}\right)}{v(R_{\rm inj})}\left(\frac{R_{\mathrm{inj}}}{R_{\mathrm{Sch}}}\right)^{\frac{5}{2}}\left(\frac{p_{\mathrm{inj}}}{p}\right)^{4} \mathrm{~d} R_{\mathrm{inj}},
\end{equation}
where we choose accretion radius $r_{\rm acc}=2GM/v^2_f\sim 0.04$ pc for the typical galactic wind velocity $v_f \simeq 500-700~{\rm km}~{\rm s}^{-1}$. $p_{\rm inj}\equiv p_{\rm inj}\left(R_{\rm inj}; r, p\right)$ is the injection momentum of an electron/positron is injected at $R_{\rm inj}~(\geq r)$ and arriving at $r$ with momentum $p$. The adiabatic compression in the advection process leads to velocity field of the accretion flow, with the momentum gain rate $\dot{p}_{\rm{ad}}$. The associated characteristic curves of $p_{\rm inj}\left(R_{\rm inj}; r, p\right)$ are obtained by solving following differential equation~\cite{Aloisio:2004hy,Lacroix:2016qpq}:
\begin{equation}\label{momentum gain}
\frac{d p}{d r}=\frac{\dot{p}_{\rm ad} (r, p)+\dot{p} (r, p)}{v(r)}\simeq \frac{\dot{p}_{\rm ad} (r, p)+\dot{p}_{\rm syn} (r, p)}{v(r)}.
\end{equation}
In above equation, approximation is due to the fact that the total radiative momentum loss $\dot{p} (r, p)$, is dominated by synchrotron radiation, and
\begin{align}
\dot{p}_{\rm ad} (r, p) &=-\frac{1}{3r^2}p\frac{\partial}{\partial r}\left[r^2 v(r)\right], \nonumber\\
\dot{p}_{\rm s y n} (r, p)&=\frac{4}{3} \sigma_{T} \frac{B^2(r)}{8 \pi} \frac{E p}{m_e^2 c^3}.
\end{align}
Following method introduced in Refs.~\cite{Aloisio:2004hy,Lacroix:2016qpq}, for a homogeneous magnetic field in the galactic halo, the analytical solution of $p_{\mathrm{inj}}\left(R_{\mathrm{inj}} ; r, p\right)$ for Eq.~(\ref{momentum gain}) is given by
\begin{equation}\label{injection_momentum}
    p_{\mathrm{inj}}\left(R_{\mathrm{inj}} ; r, p\right)=p\left[\frac{k_{0} R_{\mathrm{Sch}}^{-\frac{1}{2}}}{c} R_{\mathrm{inj}}^{\frac{3}{2}} p\left(\frac{r}{R_{\mathrm{inj}}}-1\right)+\left(\frac{R_{\mathrm{inj}}}{r}\right)^{\frac{1}{2}}\right]^{-1}.
\end{equation}
It should be emphasize that the denominator in Eq.~(\ref{injection_momentum}) may vanish and even become negative, thus leads to nonphysical values for injection momentum. To be specific, this depends on the efficiency of the accretion flow and characterizes the region of the injection parameters $(R_{\rm inj}, p_{\rm inj})$ for a given arrival point $(r, p)$. In practice, $p_{\rm inj}$ holds positive for $R_{\rm inj} < R^0_{\rm inj}$ with
\begin{equation}\label{Rinj0}
R_{\mathrm{inj}}^{0}=r+\frac{c}{k_{0} p}\left(\frac{r}{R_{\mathrm{Sch}}}\right)^{-\frac{1}{2}}.
\end{equation}
We then use the value of ${\rm min}[r_{\rm acc}, R^0_{\rm inj}]$ as an effective upper bound for the integration in Eq.(\ref{propagation_solution}). While for lower bound for the integration, we take $r=r_{\rm ISCO}$ with $r_{\rm ISCO}$ being the radius of the Innermost Stable Circular Orbit (ISCO), which indicates the innermost radius to maintain the stable orbit in the equatorial plane for a time-like test particle under small perturbations (see Appendix for details).

\subsection{Synchrotron flux density due to DM annihilations}

The power of synchrotron emission per unit frequency $\nu$ emitted by an electron of given energy $E$ and pitch angle $\theta$, in a magnetic field $B$, is written as~\cite{Ghisellini_2013}
\begin{align}
P_{\mathrm{syn}}(\nu, E_e, B, \theta) &=\frac{\sqrt{3} e^{3} B \sin \theta}{m_{\mathrm{e}} c^{2}} F\left(\nu/\nu_{c}\right), \nonumber\\
F(\nu/\nu_c) &\equiv \frac{\nu}{\nu_c}\int_{\nu/\nu_c}^{\infty} d y K_{5 / 3}(y).
\end{align}
In above equation, $F(\nu/\nu_c)$ is the standard function depicting the spectral behavior of synchrotron radiation, $K_{5/3}$ is the modified Bessel function of order $5/3$, and critical synchrotron frequency is defined as $\nu_c=3e B E^2 \sin \theta/(4\pi m^3_e c^5)$. Taking average over pitch angles, one arrived
\begin{align}
\langle P_{\rm syn}\rangle(\nu, E, B) =\frac{1}{2}\int_{0}^{\pi} d \theta \sin \theta P_{\rm syn}(\nu, E_e, B, \theta),
\end{align}

Equipped with the $e^\pm$ energy spectrum $n_e(r,E)$ in Eq.~(\ref{number density}), the emissivity of synchrotron radiation is given by convoluting $n_e$ with the averaged power $\langle P_{\rm syn}\rangle$~\cite{Ghisellini_2013}:
\begin{align}
j_{\rm s y n}(\nu, r)=2 \int_{m_{e}}^{M_{\chi}} dE \langle {P}_{\rm s y n} \rangle(\nu, E, B) n_e(r,E),
\end{align}
where the factor 2 takes into account electrons and positrons. The integrated flux density measured by a detector can be estimated as
\begin{equation}\label{flux density}
S_{\rm syn}(\nu)= \int d\Omega_{\rm obs}\int_{l.o.s}dI_{\rm syn}(\nu),
\end{equation}
where $d\Omega_{\rm obs}$ is the element of the solid angle subtended by the image plane, $dI_{\rm syn}(\nu)$ is the differential specific intensity of synchrotron emission, and the integration is performed along the line of sight. The specific intensity $I_{\rm syn}$ and emissivity $j_{\rm s y n}$ are related by standard radiative transfer equation~\cite{Rybicki:2004hfl}:
\begin{align}
\frac{d I_{\rm syn}(\nu, s)}{d s}=-\alpha(\nu, s) I_{\rm syn}(\nu, s)+\frac{j_{\rm syn}(\nu, s)}{4 \pi},
\end{align}
where the increment $ds$ along a line of sight, and $\alpha(\nu, s)$ is the absorption coefficient. The absorption process is dominated by the synchrotron self-absorption,
which is expected only to be important at low frequencies. As pointed out in Ref.~\cite{Aloisio:2004hy}, the synchrotron self-absorption is negligible if only the synchrotron losses are considered which is the same for our case. Then Eq.~(\ref{flux density}) reduces to $I_{\rm syn}(\nu)=\int ds j_{\rm syn}(\nu, s)/4 \pi$.

We next briefly discuss the issue of gravitational redshift. The synchrotron emission produced at the innermost spike region will be influenced by the gravitational potential, the modification of photon frequency due to relativistic Doppler effect and gravitational redshift must be taken into account. For this purpose, notice that quantity $I_{\rm syn}(\nu)/\nu^3$ is frame-invariance due to the Liouville theorem, which builds connection between the observed and the emitted specific intensity as
\begin{equation}\label{redsfhit}
I_{\rm obs}\left(\nu_{\rm obs}\right) = \left(\frac{\nu_{\rm obs}}{\nu_{\rm em}}\right)^3I_{\rm em}\left(\nu_{\rm em}\right)= g^3 I_{\rm em}\left(\nu_{\rm em}\right),
\end{equation}
where $g=\nu_{\rm obs}/\nu_{\rm em}$ is the redshift factor which encode both relativistic Doppler effects and gravitational redshift (see Appendix for detailed calculation). In principle, the realistic astrophysical black hole should be straightforward to extent type which is parameterized by its mass $M$ and the dimensionaless spin parameter $a=Jc^2/M$. Since current EHT resolution still can not fully determine the spin parameter of SMBH~\cite{Akiyama:2019cqa, Akiyama:2019fyp}, we here simply consider Schwarzschild case to derive a conservative limit, and it is straightforward to extent our discussion to Kerr one follow from the expression of redshift factor in Appendix. The redshift factor for Schwarzschild black hole takes a simple form as $g=\sqrt{1-R_{\rm Sch}/r}$. With this expression, substituting Eq.~(\ref{redsfhit}) into Eq.~(\ref{flux density}), one obtains observed total flux density. Here the integration of $d\Omega_{\rm obs}$ is performed over the EHT image field.

\subsection{Constraints on DM parameters from the EHT observations}\label{results and discussion}

In this work we use the observational data given in Refs.~\cite{Akiyama:2021qum, Akiyama:2021tfw}, which adopted array of the EHT to map out M87$^\star$ at a resolutions of $\sim 120~\mu$as, corresponding to a spatial resolution of $\sim 8 R_{\rm Sch}$. And the Schwarzschild radius of SMBH is estimated as $R_{\rm{Sch}}=5.9\times 10^{-4}~{\rm pc}$, corresponding to angular size $7.3~\mu {\rm as}$. And the total flux density in the image field of 120 $\mu {\rm as}$ is constrained to be 0.6~Jy and distance to us is about 16.7~Mpc. We follow the EHT collaboration to set the magnetic field strength surrounding M87$^\star$ to be $B \sim 1-30~\mathrm{G}$~\cite{Akiyama:2021tfw}, which is estimated using the reconstruction of Faraday rotation as well as the image's brightness and flux density based on a one-zone emission model. Higher values of the magnetic field predict too small electron temperature to explain the brightness temperature in the EHT image and are thus disfavored by observations. The estimate of the magnetic field strength is expected to hold at least up to $6R_{\rm{Sch}}$, which fully covers the integrated range in our calculation.

%In addition, the magnetic field strength of M87$^\star$ detected by EHT is $B \sim 1-30~\mathrm{G}$ ~\cite{ Akiyama:2021tfw}.
%The detecting method is observing linear polarization at different places with high resolution, and resolve magnetic field structure by internal Faraday rotation on small scales.
%Thus, we could assume the magnetic field region with advection and synchrotron radiation is $1-30~G$ ~\cite{Johnson:2015iwg}.

\begin{figure}
\centering
\includegraphics[width=0.8\textwidth]{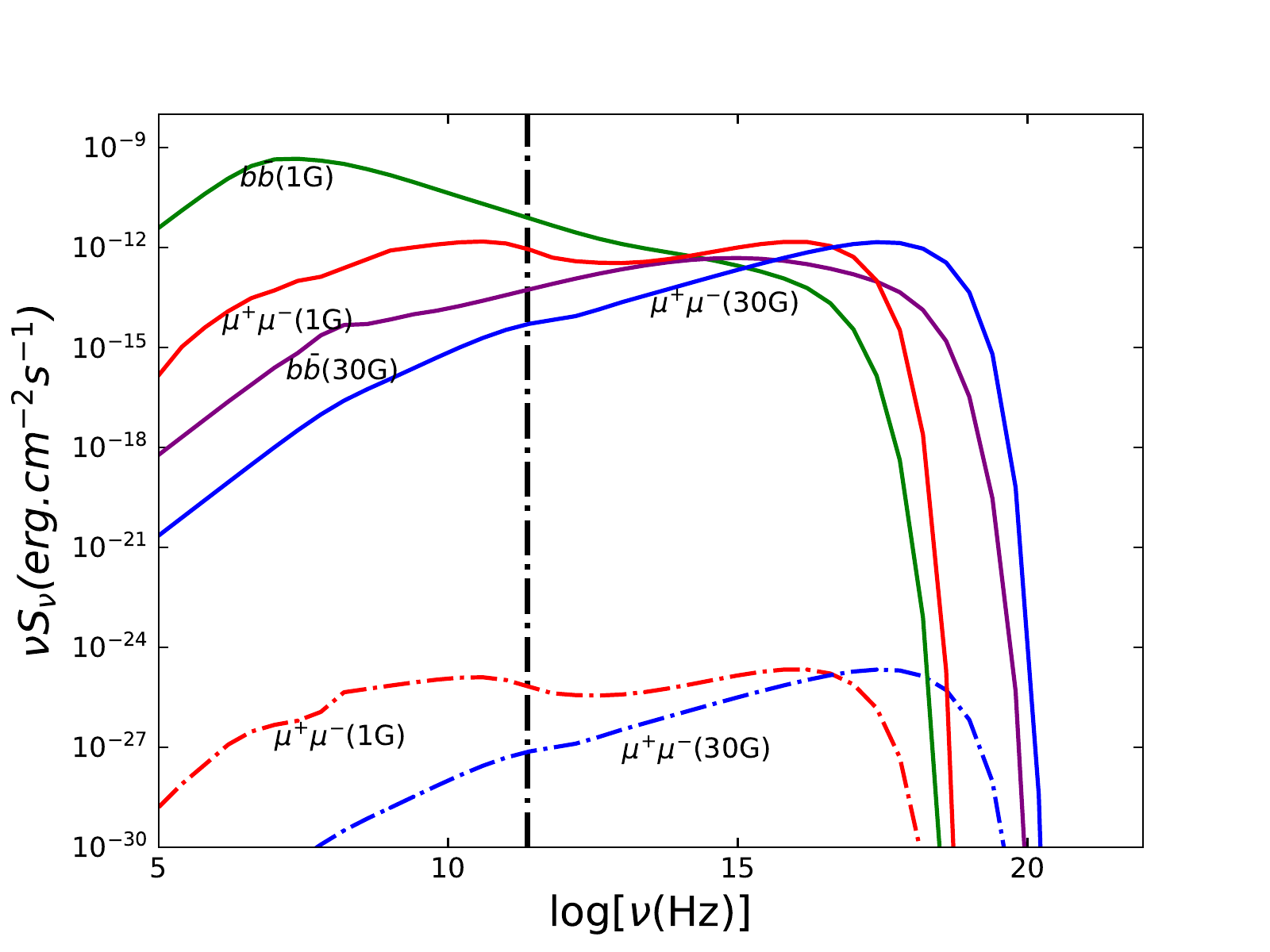}
\caption{Spectra of synchrotron emission from DM annihilation, for $b\bar{b}$ and $\mu^{+}\mu^{-}$ channels and magnetic field of 1~G and 30~G (solid lines). The DM mass and cross section are chosen as $m_{\chi}=100$~GeV and $\langle\sigma v\rangle = 10^{-30}~ \mathrm{cm}^{3}~\mathrm{~s}^{-1}$. For comparison, the contribution of regular NFW profile for $\mu^{+}\mu^{-}$ channel are also shown (dot-dashed lines).
% The solid and dashed lines show the results with magnetic fields of 1 G and 30 G. The green lines in right panel is predicted by NFW profile.
The vertical dot-dashed  line corresponds to the observational frequency, i.e., 230 GHz.}
\label{sed}
\end{figure}

%%% modified by SZQ
The spectral energy distributions (SEDs) of the synchrotron emission from the DM annihilation around the SMBH can be derived, as shown in Fig.~\ref{sed} for some examples. As a comparison, we also plot the SEDs from the NFW halo profile, which are much lower than those from the spike profile. This can also be expected according to the density profile shown in Fig.~\ref{Fig.spike}. Two processes shape the number density of electrons: the energy gain through the adiabatic compression and the energy loss through the synchrotron emission.
At high energy end, the energy loss plays an important role and therefore the radiation spectrum is contributed dominantly by the electrons from annihilation directly. However, the low-energy electrons can acquire some energy when they fall into the black hole before the gain and loss are balanced, so the SED, in the low energy range, would be harder than that without convection.
Since there are more low-energy electrons in the $b\bar{b}$ annihilation channel than those in the leptonic channels such as $\mu^+\mu^-$, the synchrotron flux at 230~GHz, at which the energy gain through the adiabatic compression would be more important, in the former case would be larger, as illustrated in Fig.~\ref{sed}. And when the magnetic field is weaker, the density can be contributed by the electrons generated in a larger space as indicated by the upper bound in Eq.~(\ref{Rinj0}) and more low-energy electrons could gain the energy during the infall as well. This larger number density of electrons would compete with the low emissivity per electron in small magnetic field and a larger synchrotron flux would be produced at 230~GHz. Therefore, the predicted flux become stronger for weaker magnetic fields, as shown in Fig.~\ref{sed}.

\begin{figure*}[htbp]
\centering
\includegraphics[width=0.49\textwidth]{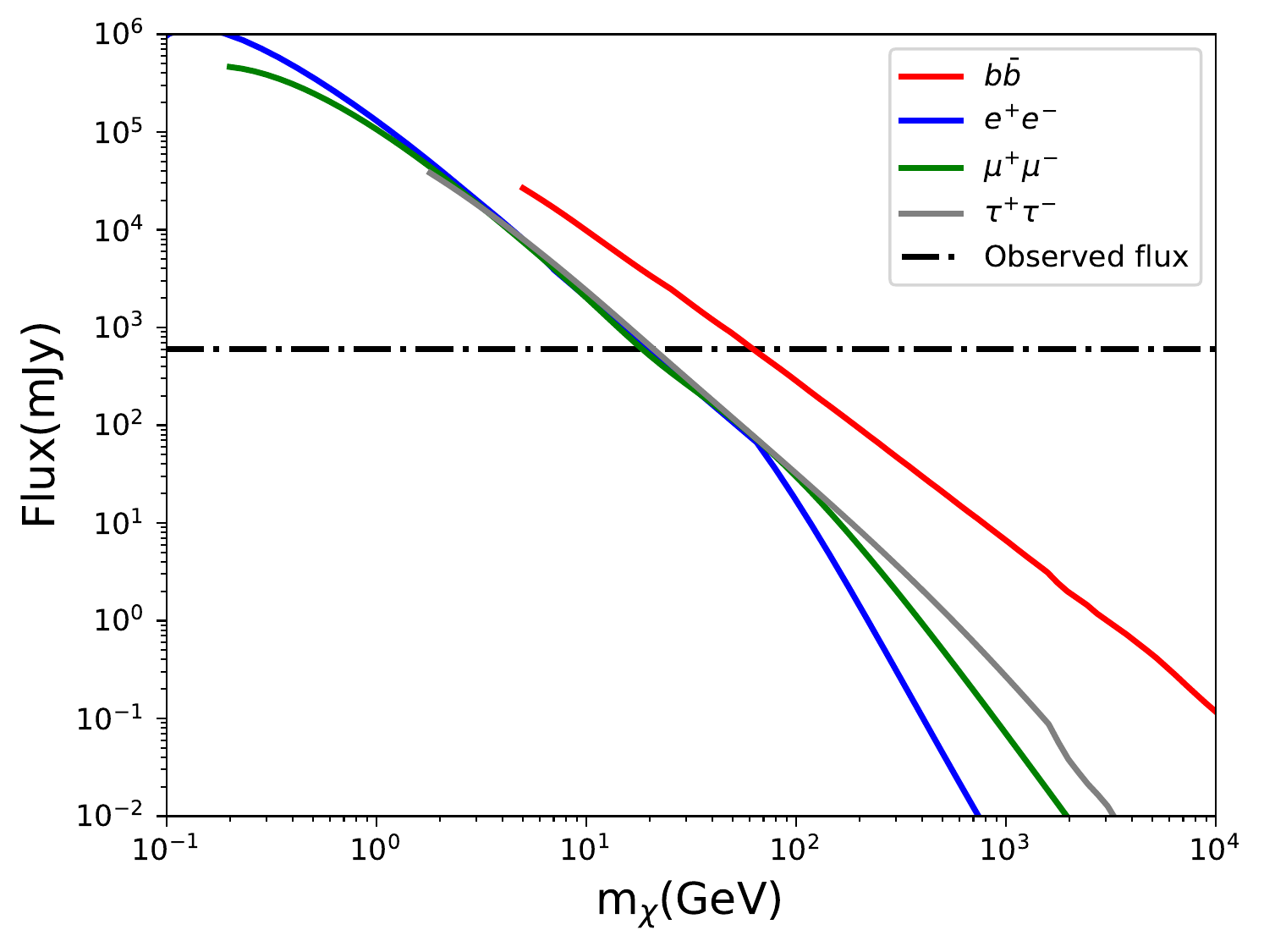}
\includegraphics[width=0.49\textwidth]{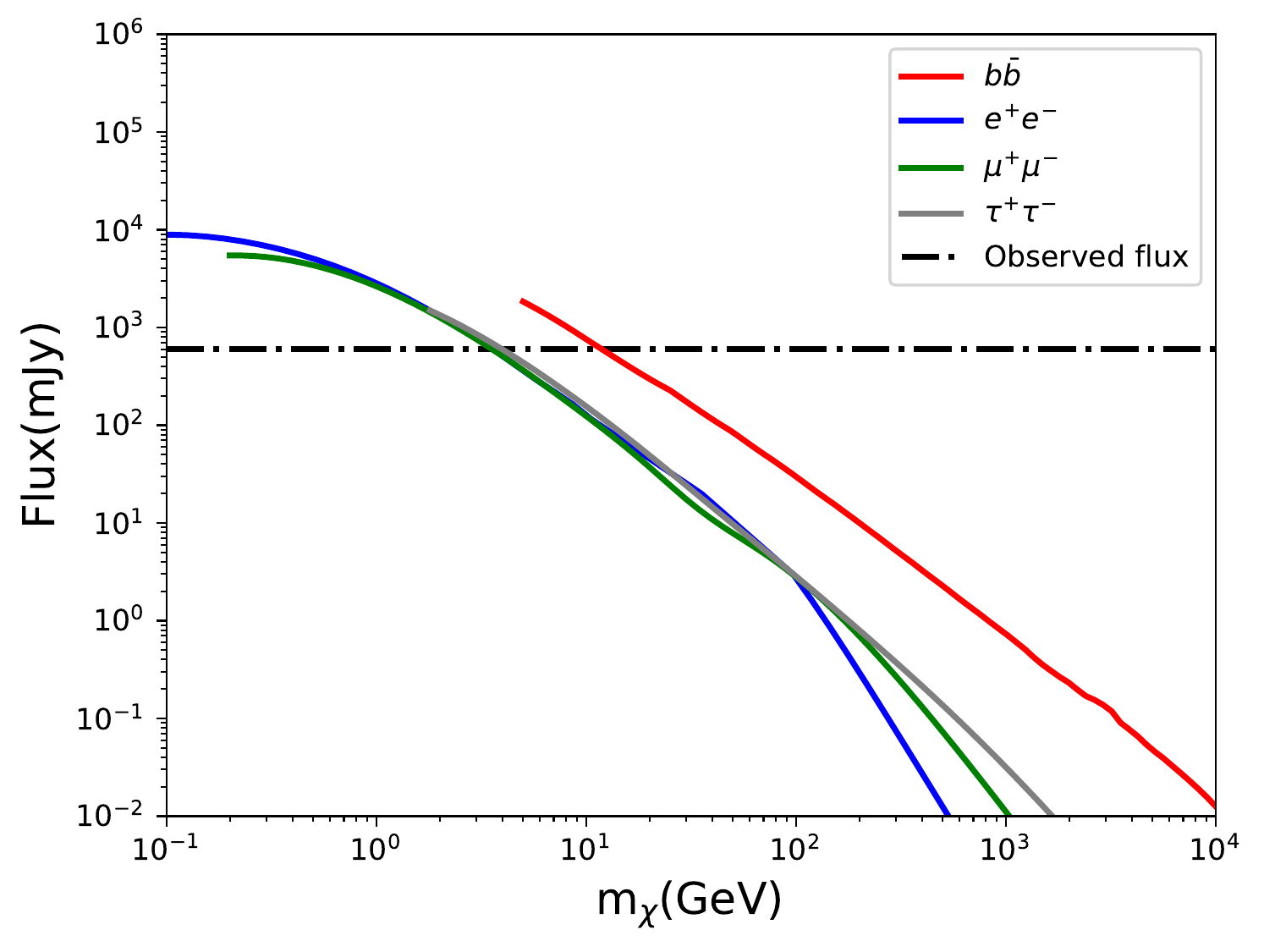}
\caption{The predicted flux of different cross section for four channels. The left/right panels are fluxes predicted by magnetic field at 1~G/30~G, respectively. And the cross-section used in left panel and right panel are $\langle\sigma v\rangle = 10^{-32}~ \mathrm{cm}^{3}~\mathrm{~s}^{-1}$ and $\langle\sigma v\rangle = 10^{-30}~\mathrm{cm}^{3} \mathrm{~s}^{-1}$. The red line is $b\bar{b}$, red line is $e^{+}e^{-}$, green line is $\mu^{+}\mu^{-}$, gray line is $\tau^{+}\tau^{-}$, and black dot-dashed line is the observed flux.}
\label{flux}
\end{figure*}

\begin{figure*}[htbp]
\centering
\includegraphics[width=0.49\textwidth]{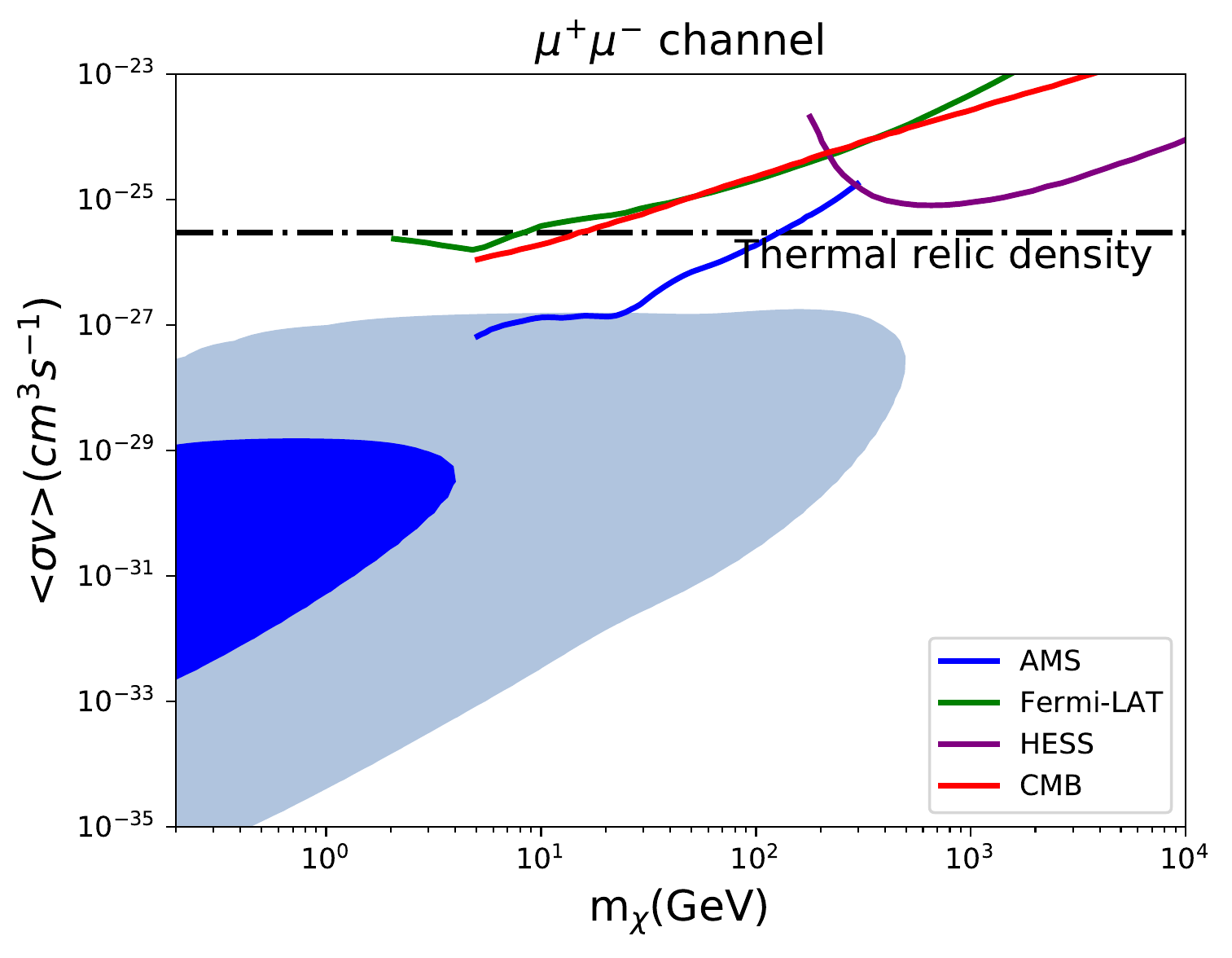}
\includegraphics[width=0.49\textwidth]{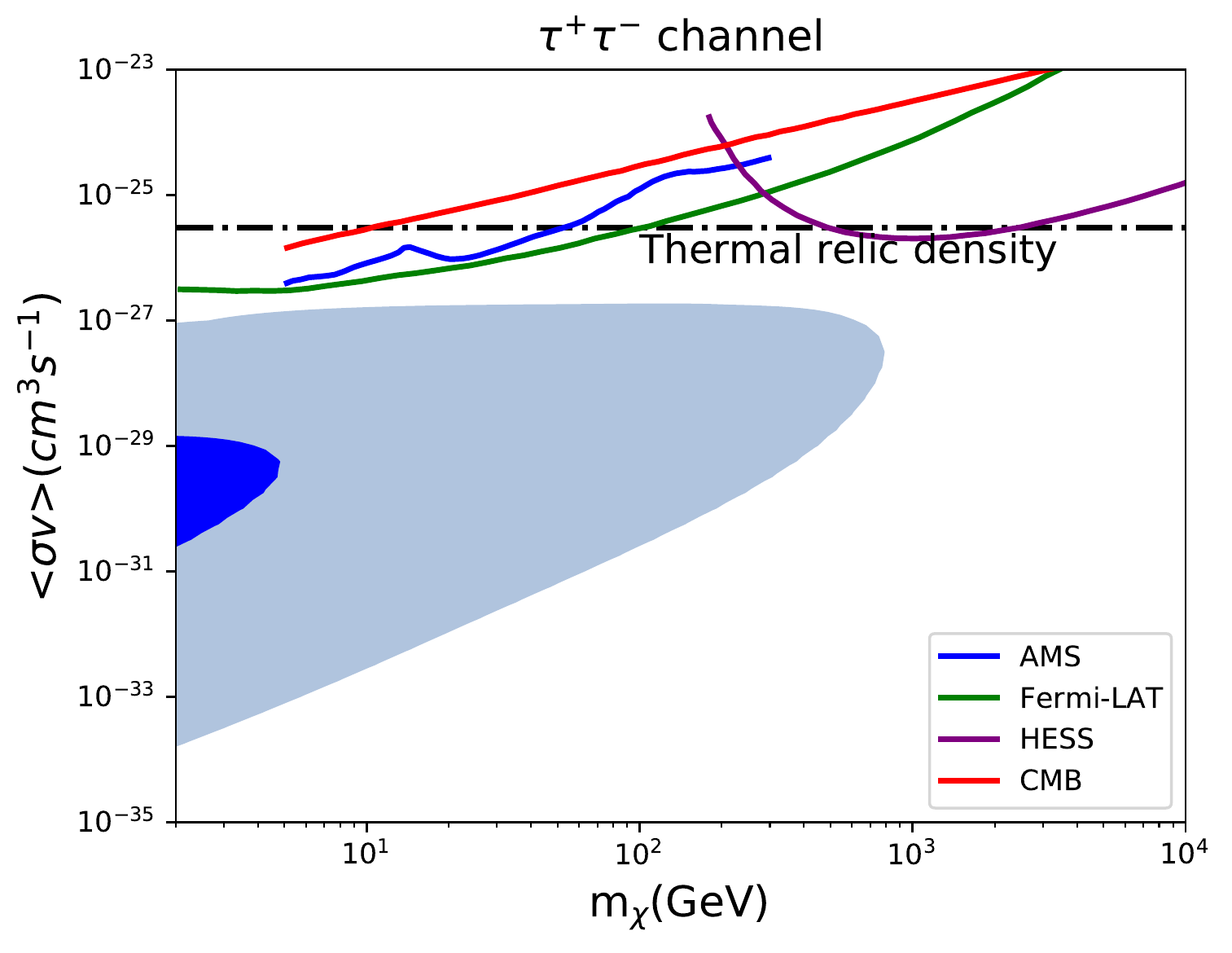}
\includegraphics[width=0.49\textwidth]{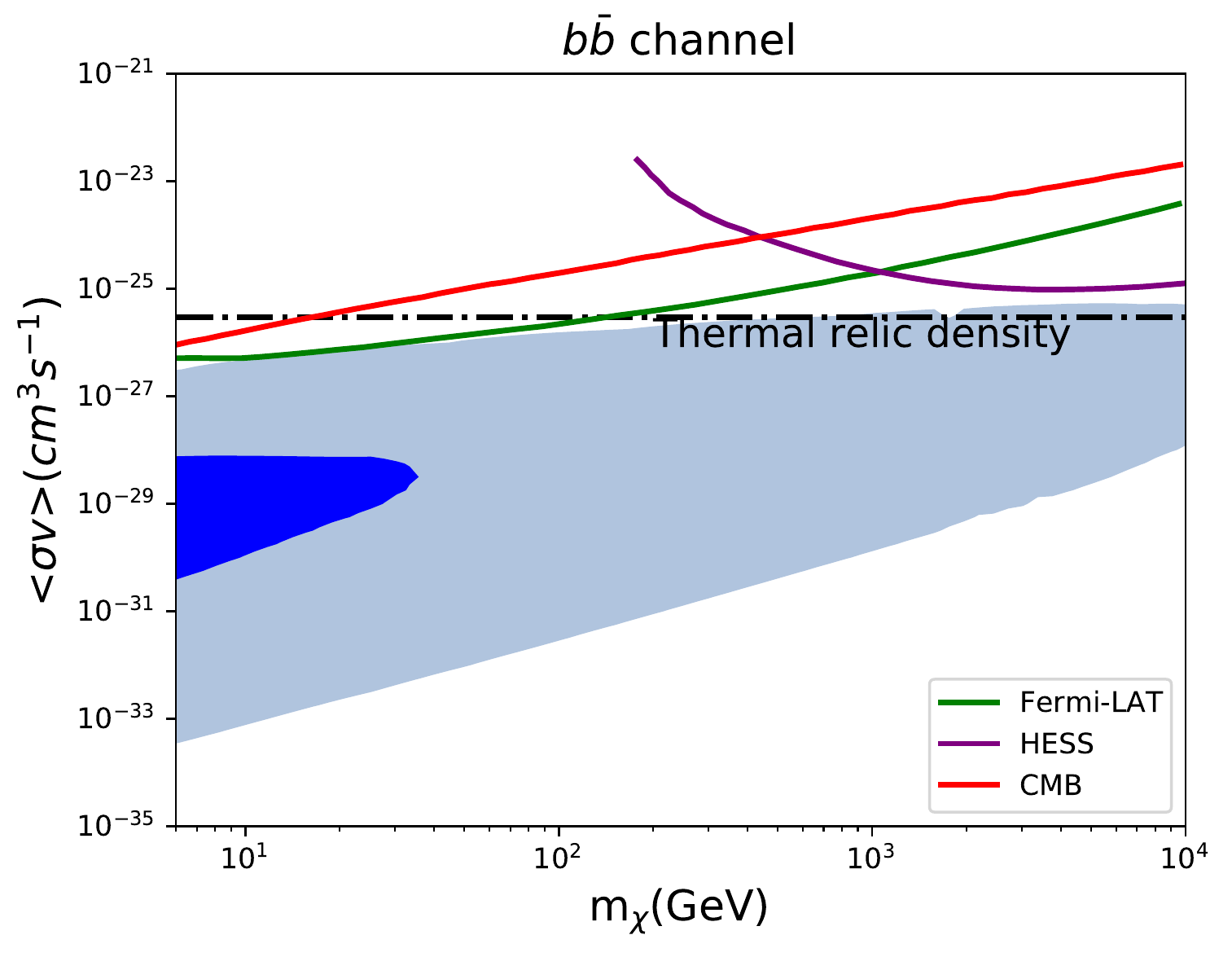}
\includegraphics[width=0.49\textwidth]{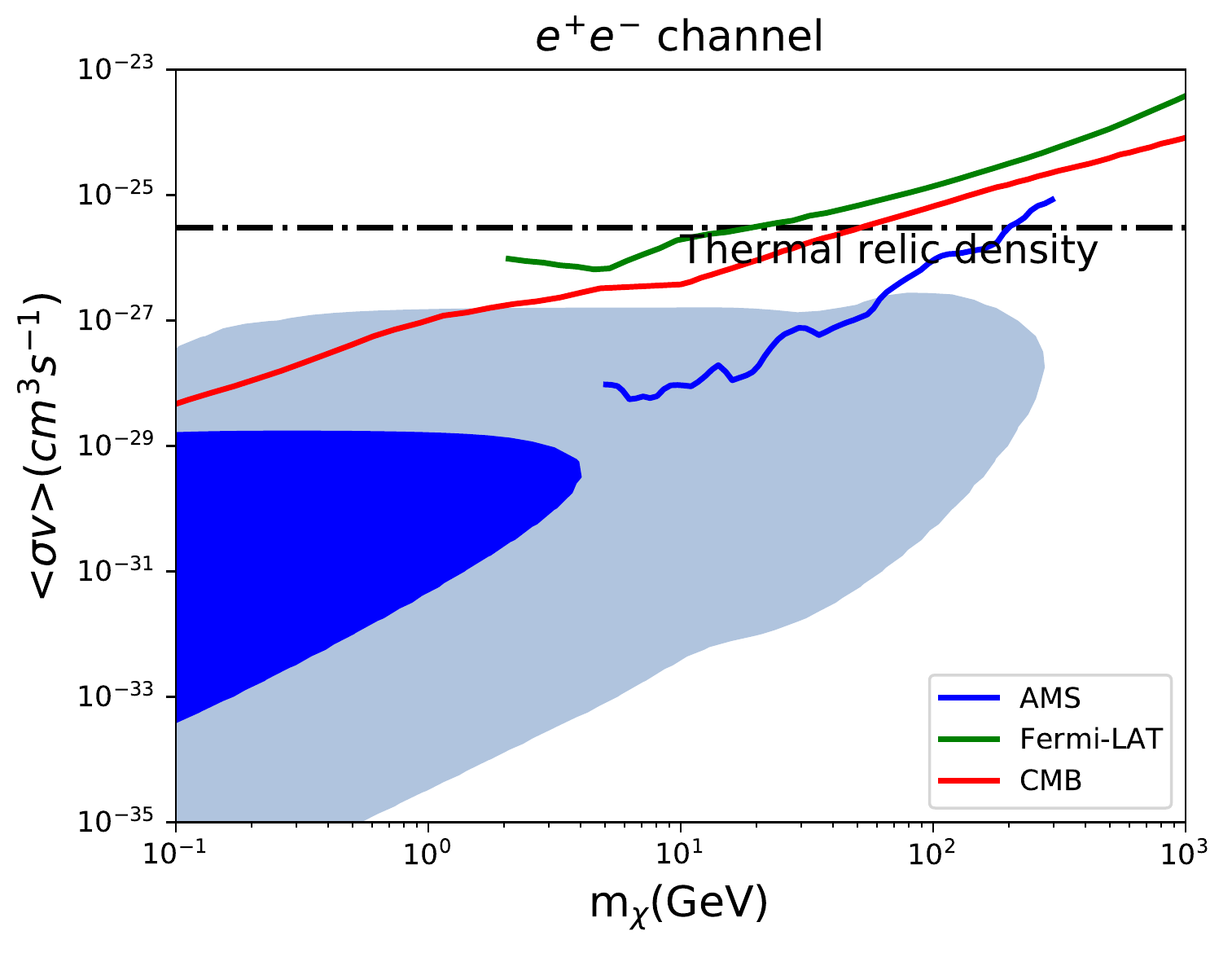}
\caption{Limits on Dark Matter Annihilation. Constraints on the WIMP pair annihilation cross section as a function of the WIMP mass at $\mu^{+}\mu^{-}$, $\tau^{+}\tau^{-}$, $b\bar{b}$, $e^{+}e^{-}$ channels by EHT M87$^\star$ result, the dark and light blue contour limits assume magnetic field are 30~G and 1~G, respectively. Other constraints, as the legend applies to all panels, from the AMS (blue line; \cite{Bergstrom:2013jra}), the Fermi-LAT (green line; \cite{Ackermann:2015zua}), H.E.S.S. (purple line; \cite{Abdallah:2016ygi}), CMB (red line; \cite{Slatyer:2015jla}) and thermal relic abundance (black dot-dashed).}
\label{bounds}
\end{figure*}

We can then set excluded regions on the $\langle\sigma v\rangle-m_\chi$ plane by comparing the expected flux with the observed flux around M87$^\star$, which by presented as the black dot-dashed line in Fig.~\ref{flux}. Since the backgrounds around the SMBH are extremely complicated, we only set conservative limits by requiring the expected fluxes not to exceed the observed one. We performed a grid scan in the $\langle\sigma v\rangle-m_\chi$ plane and calculate the exclusion regions for DM masses from sub-GeV to TeV. The results are presented in Fig.~\ref{bounds}, where the light and dark blue regions respectively correspond to limits based on the magnetic fields of 1~G and 30~G.
It is interesting to note that, the constraints for 1~G magnetic field are stronger than those for 30~G magnetic field. This is mainly because the DM-induced electron density is effectively higher due to the infall of electrons from a larger space volume for a smaller cooling in a weaker magnetic field.
For comparison, we also plot the 95\% upper limits from the AMS (blue line; \cite{Bergstrom:2013jra}), the Fermi-LAT (green line; \cite{Ackermann:2015zua}), H.E.S.S. (purple line; \cite{Abdallah:2016ygi}), and the energy injection into CMB plasma during the dark ages (red line; \cite{Slatyer:2015jla}).

%%% end SZQ

\section{Discussion}\label{summary}

In this paper, we investigated the DM-induced synchrotron emission in M87$^\star$. Assuming a spike profile of DM in the close vicinity of the SMBH, we derive new limits on the DM annihilation cross section using the  EHT observations at 230~GHz, for $\mu^{+}\mu^{-}$, $\tau^{+}\tau^{-}$, $b\bar{b}$ and $e^{+}e^{-}$ channels for typical mass regions of WIMP DM. Since the spike density in the inner regions would be further enhanced for relatively small annihilation cross section, the expected flux would also be amplified. As a consequence, unlike other existing limits which are usually more sensitive to the large annihilation cross sections, in our case the small annihilation cross section within $\langle\sigma v\rangle \sim 10^{-34}-10^{-27}{\rm cm}^3~{\rm s}^{-1}$ is more severely constrained. On the other hand, the spike density would be depleted for sufficiently large $\langle\sigma v\rangle$ and reaches saturation, resulting in no constraints when $\langle\sigma v\rangle$ is larger than $\sim 10^{-27}{\rm cm}^3~{\rm s}^{-1}$. Moreover, due to a non-monotonic dependence of the injection spectrum on the magnetic field, more stringent constraints are given for a lower magnetic field.

If the DM spike does exist, our results are more sensitive in the small cross section region, which are complementary to other experiments. The DM distribution without enhancement from the spiky structure predicts much lower synchrotron emission flux, and thus no effective constraint can be obtained.
%On the other hand, the EHT observations may be used to constrain the spike model itself since the spike model is still under debate~\cite{Merritt:2002vj,Bertone:2005hw,Ferrer:2017xwm,Bertone:2018krk}.
The forthcoming EHT observations of the SMBH in our Milky Way is expected to improve significantly the results of the current work due to its proximity. In addition, observations at multiple frequencies should also be very helpful in better identifying the DM signal or constraining the model parameters.
%In this direction, the combination of multi-wavelength signals is a promising way. We expect that the EHT observations in the near future with higher resolution will significantly improve the constraints on spike density profile, which may provide a unique opportunity to test the morphology of innermost density profile of DM halo.

\section*{Acknowledgements}

We thank Yifan Chen, Yi-Zhong Fan, Lei Feng, Lingyao Kong, Xu Pan, Chi Tian and Yue-Lin Sming Tsai for helpful discussion and valuable comments in various aspects. R.D. is supported in part by the National Key R\&D Program of China (2021YFC2203100). Z.Q.S is supported by the National Natural Science Foundation of China (NSFC) under Grants No. U1738210 and No. 12003074; X.H. and Q.Y. are supported by Chinese Academy of Sciences and the Program for Innovative Talents and Entrepreneur in Jiangsu.

\bibliography{references}
% \addcontentsline{toc}{section}{Bibliography}
\bibliographystyle{JHEP}

\section*{Appendix: The calculation of gravitational redshift}
In this appendix, we provide some details about the calculation of the gravitational redshift effect. The Kerr mertic is parameterized by the black hole mass $M$ and the dimensionaless spin parameter $a=Jc^2/M$, where $J$ is the angular momentum of the black hole. In the Boyer-Lindquist coordinates and convention of natural units, the Kerr metric element is written as~\cite{Boyer:1966qh}
\begin{align}\label{kerr}
\mathrm{d} s^{2}&=-\left(1-\frac{2 M r}{\Sigma}\right) \mathrm{d} t^{2}-\frac{4 a M^{2} r \sin ^{2} \theta}{\Sigma} \mathrm{d} t \mathrm{~d} \phi+\frac{\Sigma}{\Delta} \mathrm{d} r^{2} \nonumber\\
&+ \Sigma \mathrm{d} \theta^{2}+\left(r^{2}+a^{2} M^{2}+\frac{2 a^{2} M^{3} r \sin^{2} \theta}{\Sigma}\right) \sin ^{2} \theta \mathrm{d} \phi^{2}.
\end{align}
where $\Delta=r^{2}-2 M r+a^{2} M^{2}$, $\Sigma=r^{2}+a^{2} M^{2} \cos ^{2} \theta$. The event horizon correspond to the outer root of $\Delta=0$, which yields $r_{\mathrm{evt}}=M\left(1+\sqrt{1-a^{2}}\right)$. With metric element in Eq.~(\ref{kerr}), the redshift factor $g$ is defined as
\begin{equation}
g\equiv\frac{E_{\mathrm{obs}}}{E_{\mathrm{em}}}=\frac{\nu_{\mathrm{obs}}}{\nu_{\mathrm{em}}}=\frac{k_{\alpha} u_{\mathrm{obs}}^{\alpha}}{k_{\beta} u_{\mathrm{em}}^{\beta}},
\end{equation}
$k_{\alpha}$ and $k_{\beta}$ are the 4-momentum of the photon, $u^{\alpha}_{\rm obs} = (-1,0,0,0)$ is the 4-velocity of the distant observer, and $u^{\alpha}_{\rm em} = (u^{t}_{\rm em},0,0,\Omega u^{t}_{\rm em})$ is the 4-velocity of the emitter. The angular frequency of a test-particle at the emission radius $r_e$, $\Omega = u^{\phi}/u^{t}$, in the free-fall model, can be written as~\cite{Pu:2016qak},
\begin{equation}
\Omega(r, \theta)=\frac{2a^{2} M^{3} r}{(r^2 +a^2)^2 -a^2(r^2-2Mr+a^2)\sin^2\theta}.
\end{equation}
The source emission is assumed to be monochromatic and isotropic, with a Gaussian intensity. Using the normalization condition $g_{\mu\nu}u^{\mu}_{\rm em}u^{\nu}_{\rm em}=-1$, we have
\begin{equation}
u_{\mathrm{em}}^{t}=-\frac{1}{\sqrt{-g_{t t}-2 g_{t \phi} \Omega-g_{\phi \phi} \Omega^{2}}},
\end{equation}
and therefore,
\begin{equation}\label{redshift}
g=\frac{\sqrt{-g_{t t}-2 g_{t \phi} \Omega-g_{\phi \phi} \Omega^{2}}}{1+\lambda \Omega}.
\end{equation}
Here, $\lambda = k_{\phi}/k_{t}$ is a constant of the motion along the photon path. For Schwarzschild black hole with $a=0$, Eq.~(\ref{redshift}) simplified to $g=\sqrt{1-R_{\rm Sch}/r}$.

For Kerr black hole, the radius of ISCO has expression~\cite{Reynolds:2013rva}:
\begin{align}
r_{\rm ISCO}=M\left(3+C_{2} \mp \sqrt{\left(3-C_{1}\right)\left(3+C_{1}+2 C_{2}\right)}\right),
\end{align}
where the $\mp$ sign is for test particles in prograde and retrograde orbits, respectively, relative
to the spin of the BH and we the coefficients $C_{1,2}$ are given as
\begin{align}
C_{1}&=1+\left(1-a^{2}\right)^{1 / 3}\left[(1+a)^{1 / 3}+(1-a)^{1 / 3}\right], \\
C_{2}&=\sqrt{3 a^{2}+C_{1}^{2}}.
\end{align}
Again, taking $a=0$, one obtains $r_{\rm ISCO}=3R_{\rm Sch}$ for Schwarzschild black hole.
\end{document}